# Ice Dome Construction for Large Scale Habitats on Atmosphereless Bodies

Stefan Harsan Farr

shfarr@gmail.com

## Abstract

One of the greatest difficulties that space exploration faces is the lack of technology necessary to establish large volumes of habitable spaces on site. Both transporting the pre-built enclosures or transporting the equipment necessary for building them on site from conventional materials face the same enormous problem: the need to transport huge quantities of material into space, which is technically close to impossible. The current paper, explores the possibility and one approach of building these large spaces from an alternative material, water ice, a material that is a prerequisite for any settlement. The feasibility of dome shaped, pressurized, water ice buildings is analyzed from a structural integrity point of view and the possibility of building them with a technique using water sublimation and deposition onto a thin plastic film, a process which requires extremely little construction equipment with respect to the resulting habitable space.

# 1  Introduction

There is great consensus about the the fact that space exploration is a necessary step in humanity's evolution, the big dissension lies in it's urgency. Scientists and space fans, myself included, see it as imperative urgent step that humanity must take as soon as possible no matter the cost, but policy makers greatly disagree. These people, who in fact hold the funding in their hands, tend to be more cautious and conservative and are less inclined to redirect large amounts into such new and uncertain directions. Unfortunately, space exploration lies in a sector which requires funds vastly beyond the amount any policy maker would be willing to gamble.

The alternative, as so many times along the history, is the private sector which is willing to invest a great deal but with a catch: the investment needs to become not only feasible but profitable, in relatively short time, a prerequisite for surviving in the private sector. For this to happen, space exploration must go beyond sending a few people to these distant destinations, it needs to establish colonies of hundreds if not thousands of people. Profitability needs volume, which brings costs down and raises new challenges creating dynamics that drive themselves ultimately into profitability.

Space however, is an inhospitable destination. To establish, colonies with a large degree of self sustainability, a necessary condition imposed by the difficulties in transportation, colonies will need first of all, large habitable enclosures that are suitable for working, growing food and living in general. People need a certain degree of dexterity to be able to efficiently solve problems, therefore cumbersome space suits and small living enclosures will not be adequate for large populations.

Mainstream approaches such as building structures on site from conventional materials like rock, digging underground structures or using inflatable compartments are impractical for large structures due to either the necessary equipment or the necessary material, both of which need to be transported from Earth. Even though transportation costs have dropped considerably in the recent years, transporting such large volumes is still highly impractical.

The proposed approach, uses an alternative material, water ice, the presence of which is a must for any colony and takes advantage of the inhospitable conditions that exist at these destination to execute the construction with very little equipment that needs to be transported from Earth.

# 2  Implementation Considerations

The method aids the construction of large to very large (ex: from 900 m$^2$ / 6,500 m$^3$ to 30,000 m$^2$ / 143,000 m$^3$) dome shaped structures made of water ice, which would allow partial or complete Earth like atmospheric pressure underneath and temperatures ranging from minus a few negative degrees Celsius to positive degrees Celsius, depending on the size, thickness and the possibility for internal thermal insulation of the structure.

Preliminary calculations, however, show that the theoretical size of these structures is not limited to these values. Structures of almost any size can be constructed using this method and material, but special conditions will be imposed in order to maintain the balance of forces.

The construction method requires special environmental conditions, therefore it may not be suitable under certain conditions. On Earth for example, it can only be applied under special laboratory conditions but it can be suitable for places like the Moon, because it relies on the characteristics of the hostile environment that exists there.

## 2.1  Principles Of The Construction Method

The method consists in the deposition of water vapor directly onto the surface of an inflated, dome shaped, ETFE (Ethylene tetrafluoroethylene, see 2.3.1) film the exterior of which is exposed to the vacuum of space or a very low temperature, Fig. 1. Over a period of time the water will accumulate in an ever thicker layer of solid ice, which then can be pressurized to habitable pressures and the temperature can be brought to tolerable or comfortable values.

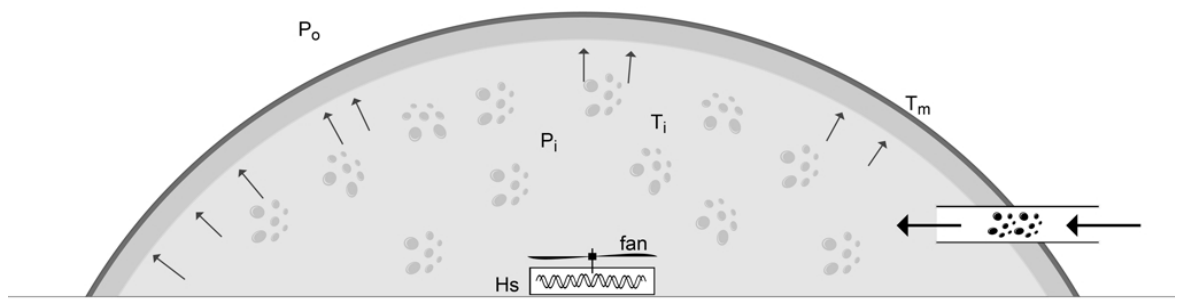

Fig. 1: Cross section of a dome shaped building during construction. $T_m$ – membrane temperature, $T_i$ – inside temperature, $P_i$ – inside pressure, $P_o$ – pressure outside, $H_s$ – heat source, fan – fan to encourage upward air flow

The area where the dome will lie needs to be cleared of debris to prevent puncturing of the film and needs to be very flat to ensure equal distribution of forces in the dome.

The ETFE dome, which is prefabricated on Earth and transported on site needs to be inflated first with an inert gas like Nitrogen till it reaches its desired shape. This step is important to prevent the formation of the ice before the dome is formed.

After the dome is filled, water is introduced into the dome via an entry point and depending on the form it is introduced it will evaporate or sublimate, due to the low pressure inside the dome. This process imposes the condition that the pressure inside the inflated ETFE dome be less than  600 Pa (Fig. 3). Also, for the method to work, the temperature inside the dome must be larger than the deposition temperature of the water, at the operating pressure, but low enough not to melt the ice forming on the contact zone, which is thermally exposed to the outside loosing heat and therefore providing a surface for deposition.

The inside temperature and pressure will vary on the local atmospheric pressure. Lower temperatures are better, because they facilitate formation of harder frost (water deposition), ideally under -18°C. Theoretically, pressure would depend on the gravitational acceleration on the surface which affects the weight of the dome and hence the pressure needed to keep it inflated, however, due to the light nature of the ETFE foil, this latter value would have very little influence and can be neglected altogether (See 2.9 Case Study: the inflating pressure is $P_i = 7.2*10^{-3}$ Pa), unless conditions are made critical by higher local atmospheric pressure which gives very little margin for ice deposition (for example on Mars).

If the temperature inside the dome falls below deposition temperature the vapor will turn into snow and fall onto the ground where it can be heated and sublimated again.

In spite of the magnitude of the construction, this method will not need any heavy equipment. By varying the water input to keep optimal pressure, and heating ice on the bottom of dome, the dome will construct itself over time. When the desired thickness of the dome shell is reached, the process can be stopped, the interior can be pressurized and temperature can be regulated accordingly.

## 2.2 Geometric Aspects

For the symmetrical dome cross section in Fig. 2, the following defining values can be identified:

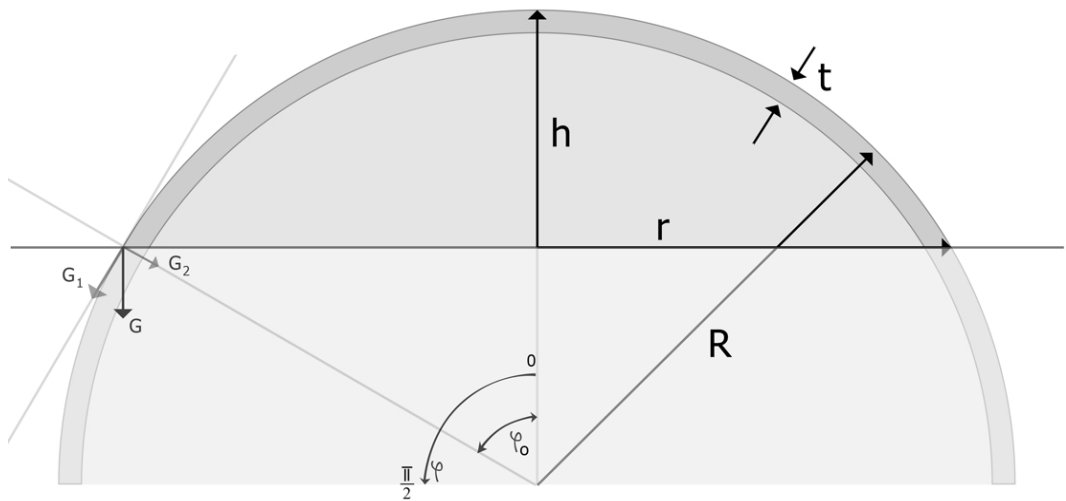

Fig. 2: Symmetrical dome cross section

- R (m), the radius of the curvature of the dome,
- h (m), the height of the dome,

- t (m), the thickness of the dome's shell,
- φ₀, the angle that defines the base of the dome, if Pi/2, the dome is a complete hemisphere.
- φ, the angle defining a point on the arc of the dome. It starts from the top and takes values in the [0, Pi/2] interval

The values important for the calculation of the parameters of the dome are given by the formulas:

$\cos(\varphi) = \frac{(R-h)}{R}$   the cosine of the angle corresponding to a point on the surface

$r(m) = \sqrt{R^2 - (R-h)^2}$   the radius of the base circle (footprint)

$A_{base}(m^2) = \pi \times r^2$   the area of the base

$V_{cal}(m^3) = \frac{\pi \times h^2}{3}(3 \times R - h)$    $A_{shell_e}(m^2) = 2 \times \pi \times R \times h$   volume and area of a spherical calotte

$A_{usable}(m^2) = \pi \times r^2 - \pi \times (r-t)^2$   the interior (usable) surface and volume

$V_{usable}(m^3) = \frac{\pi \times (h-t)^2}{3}(3 \times (R-t) - (h-t))$   volume of the resulting space

$V_{shell}(m^3) = \frac{\pi \times h^2}{3}(3 \times R - h) - \frac{\pi \times (h-t)^2}{3}(3 \times (R-t) - (h-t))$   the volume of the shell itself

From these the material needs can be calculated:

$M_{dome}(Kg) = m_s \times V_{shell}$   the mass of water needed for the construction of the shell

$A_{ETFE}(m^2) = A_{shell_e} + A_{base}$   the total area that needs to be covered by ETFE during deposition

The total mass of ETFE foil used will depend on the thickness of material used.

$M_{ETFE}(Kg) = 0.045 * A_{ETFE}$   For the $m_{s\ foil}$ = 0.0254 mm, thick foil. This value is important,

because it is a component that needs to be transported from Earth to the construction site.

## 2.3 Assumptions And Requirements

The first requirement is that water must exist in place. This can be mitigated by the fact that no place would be chosen for long term settlement unless large amounts of water can be found nearby.

The second condition is that the ice structures must be either insulated from the exterior with reflective material from solar radiation or the solar radiation needs to be low enough not to melt the ice during exposure. On distant bodies like Callisto, Europa this is solved by their distance from the Sun but on objects closer to the sun, such as the Moon, the suitable places will be the permanently dark regions which happen to be the only places on the Moon to possibly preserve water to these days.

As a construction material water needs very specific conditions most importantly because it has a very low melting/sublimation point. The method takes advantage of the phase transition of water at low pressures and temperature where the liquid state is skipped altogether allowing transition from water vapor directly into solid.

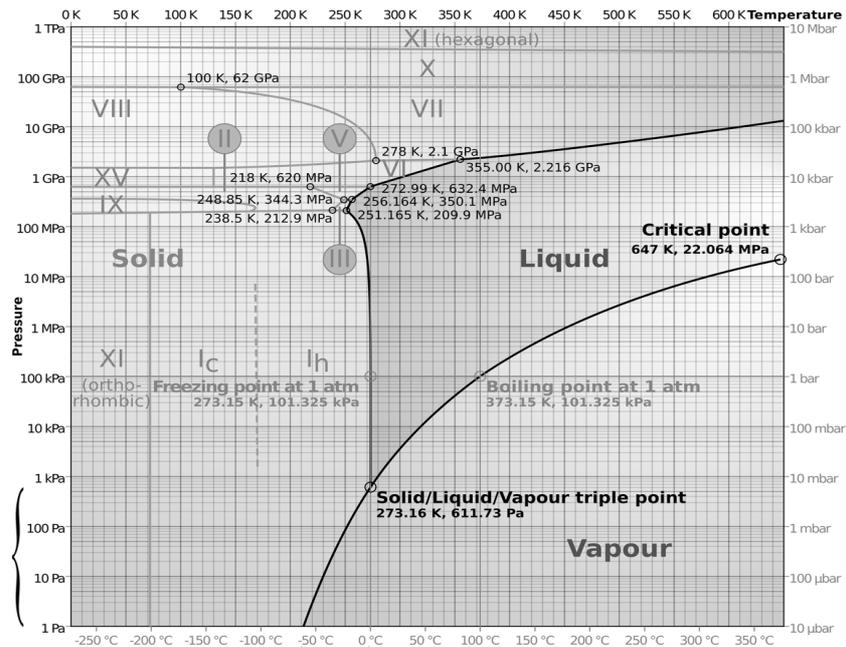

*Fig. 3: Phase diagram of water* [6]

The region of interest can be observed on the phase diagram of water (Fig. 3) and is bounded by the approximate values P = [1, 600] Pa and T = [-60, 0] °C, where the transition from vapor to solid is possible.

For the method to work, the exterior of the construction needs to lose heat to the outside either by thermal radiation or convection, which can be achieved in either very cold environments or in vacuum with very little incident thermal radiation.

$$Q = \sigma \times A_{shell} \times T^4$$

The method is not suitable for building anywhere in the solar system, but favorable conditions exist on many tantalizing destinations, such as the Moon, large asteroids, Callisto, Europa and even on Mars at high altitudes.

In the current paper ideal conditions are considered for all calculations:

- the total lack of atmosphere will lead to complete vacuum which means outside pressure will be considered to be 0 Pa.
- heat loss will be due to thermal radiation in total darkness and outside incident thermal radiation will be equal to that of the cosmic background 3 °K.
- gravitational acceleration will be considered constant over the height of the dome and equal to that on the ground zero for the bodies in case.
- the dome will have ideal shape and the water deposition constant overt the entire surface allowing equal distribution of the forces.

## 2.4 Process Consideration

### 2.4.1 Inflating Pressure

The pressure needed to counteract the weight of the ETFE dome foil is given by the formula of the force exerted by internal pressure:

$$F = P \times A$$

F is the weight of the ETFE dome

$$G_{ETFE} = m_{s\,ETFE} \times g_{local}$$

and the specific mass of the sheet is given by m² for a predefined thickness. In this case we obtain:

$$P_{ETFE} = g_{local} \times m_{s\,ETFE}(Kg/m^2) = 9.81 \times 10^{-5} \times g_{local} \times m_{s\,ETFE}(KPa)$$

It can be observe, that pressure will be the same regardless of the dome size. What matters is the the type of ETFE chosen and the place where the dome is built.

### 2.4.2 Heat Loss

In order to maintain constant temperature inside the dome and keep the gas at optimum temperature, the rate at which the dome looses heat must be compensated by a heating mechanism, like microwave or infra read radiation sources.

The radiative power, Q (it is noted with Q, not to be confused with the Pressure, which is noted with P) can be written:

$$Q(W) = \varepsilon \times \sigma \times A_{shell} \times (T_i^4 - T_o^4)$$

Where ε represents the emissivity of water, $ε_w = 0.95$ [7], and $T_o$ represents the temperature of the cosmic background, which is around $T_o = 3°K$.

If the operation temperature is chosen to be let's say 223°K (approximately -50°C), the formula can be written as:

$$Q(W) = 5.6703 \times 10^{-8} \times 223^4 \times A_{shell} = 133 * A_{shell}$$

For convenience, the heat loss can be defined per unit of area,

$$q_s(W/m^2) = 133$$

The necessity of thermal coat for the foil is due to the high emissivity of water. For large domes, the heat loss would require enormous heat generators, however, with silver coting applied to the exterior of the dome, which has an emissivity, $ε_s = 0.03$ [7], the rate of heat loss would be reduced to:

$$q_{ss}(W/m^2) = 4.2$$

This coating will continue to help insulating later on when the dome will be pressurized and have even higher interior temperature.

## 2.5 Structural Considerations

This construction method is not limited to domes. Any type of mono volume construction can be obtained by water deposition, as long as the manufacturing of the ETFE foil permits the shape, however due to the somewhat weaker mechanical properties of ice, relative to other construction materials (Just 1/5 of the crushing strength of concrete), the curved geometry of hemicylindrical or dome structures are better suited because of their efficiency in dispersing stress evenly throughout the entire structure and thus avoiding points of high stress which are the main cause of structural collapse.

Another benefit of the dome shape is that it offers the highest ratio of usable volume and area relative to building material of all shapes.

### 2.5.1 Forces Acting On The Dome

During the construction process, the major forces acting on the dome will be the weight of the dome, which is a downward force the resultants of which are channeled by the geometry and distributed throughout the shell generating compression and tensile stress. This force depends on the mass of the dome and the local gravitational acceleration constant:

$$G_{dome}(N) = M_{dome} \times g_{local}$$

Once the construction is complete and the dome is pressurized to P (KPa)= 101, pressure on Earth at sea level (one atmosphere), another force appears which is generated by the pressure inside the dome. The upward resultant of this force:

$$F_{up}(N) = P * A_{usable} \quad ,$$

will oppose the weight of the dome effectively reducing the it to a residual weight:

$$G_r(N) = G_{dome} - F_{up}$$

In the resulting construction, $G_r$ must be positive or the dome will be lifted off the ground.

### 2.5.2 Dome Stress Under It's Own Weight

Two kinds of stress act inside the dome shell, both of which are different resultants given by the specific weight of the construction material and the geometry of the dome. In the case of a symmetrical dome this stress can be calculated according to Ansel C. Ugural by the formulas:

$$\text{The meridional stress} \quad C_\varphi(KPa) = \frac{-R \times (t \times m_s \times g_{local})}{t \times (1 + \cos(\varphi))} \times \frac{1}{1000} \quad [1],$$

The 1/1000 is needed for the conversion to (KPa), as the result of the formula is in (Pa), which represents the compressive stress the maximum value of which is reached at the bottom of the dome (Fig. 4),

$$\text{and the hoop stress} \quad T_\varphi(KPa) = -\frac{R \times (t \times m_s \times g_{local})}{t} \times \left(\cos(\varphi) - \frac{1}{1 + \cos(\varphi)}\right) \times \frac{1}{1000} \quad [1],$$

which is the tensile stress, the maximum value of which is reached also at the bottom edge of the dome (Fig. 4).

In case the dome is supporting its own weight, the thickness is canceled out and does not impact the stress values in various points of the dome.

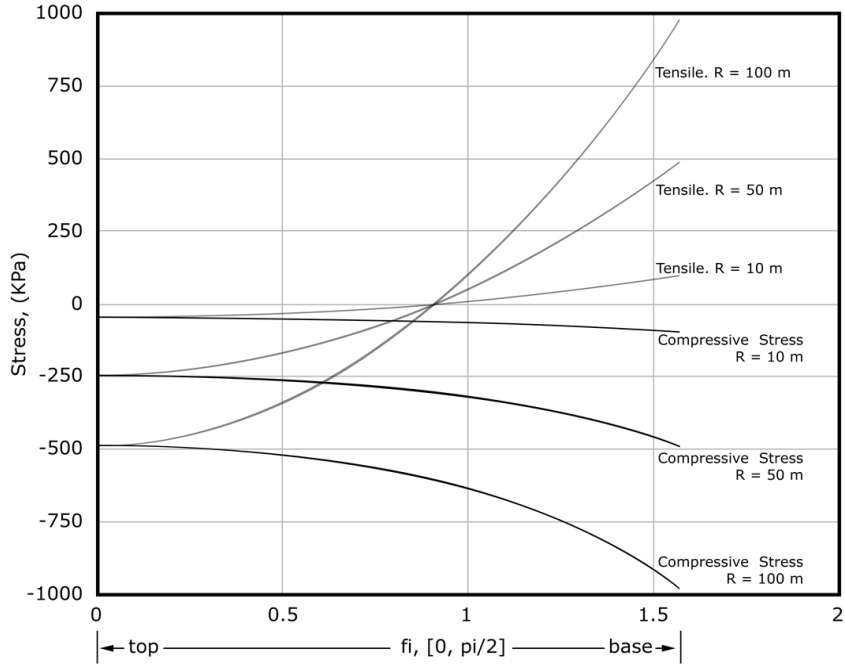

*Fig. 4: A visual representation of the values of the tensile and compressive stress for a 1 m thick dome of ice (ms = 1000 Kg/m³) supporting it's own weight on Earth (g = 9.8 m/s²), for various values of radii (R = 10 m, 50 m and 100 m respectively).*

### 2.5.3 Pressurized Dome

The force applied by the internal pressure acts perpendicularly onto the internal surface of the dome.

In case of a perfect sphere, this force would be equal in all directions and would diffuse almost exclusively into tensile stress in the surrounding material. In this case however, the dome does not have a bottom, therefore there will be an upward resultant which will act as lift force and an outward resultant which will cause tensile stress. Under these conditions, the formulas for calculating the stress in the material become:

The meridional stress $C_\varphi(KPa) = \dfrac{-R \times (t \times m_s \times g_{local} - 1000 \times P_i)}{t \times (1 + \cos(\varphi))} \times \dfrac{1}{1000}$ ,The x1000 in

$1000 \times P_i$ needs to be added by because it is defined as (KPa) and the factors in front yield (Pa).

The hoop stress $T_\varphi(KPa) = -\dfrac{R \times (t \times m_s \times g_{local} - 1000 \times P_i)}{1000 \times t} \times \left(\cos(\varphi) - \dfrac{1}{1 + \cos(\varphi)}\right) + P_i \times \sin(\varphi)$ ,

The inside pressure relieves some of the initial compression pressure and tensile stress due to the lift force which counter-acts the weight of the dome. The residual force, generated by the pressure (the part that is not discharged as lift force) will manifest almost exclusively in tensile stress.

A comparison between the evolution of the forces with and without inside pressure is presented in the Fig. 5.

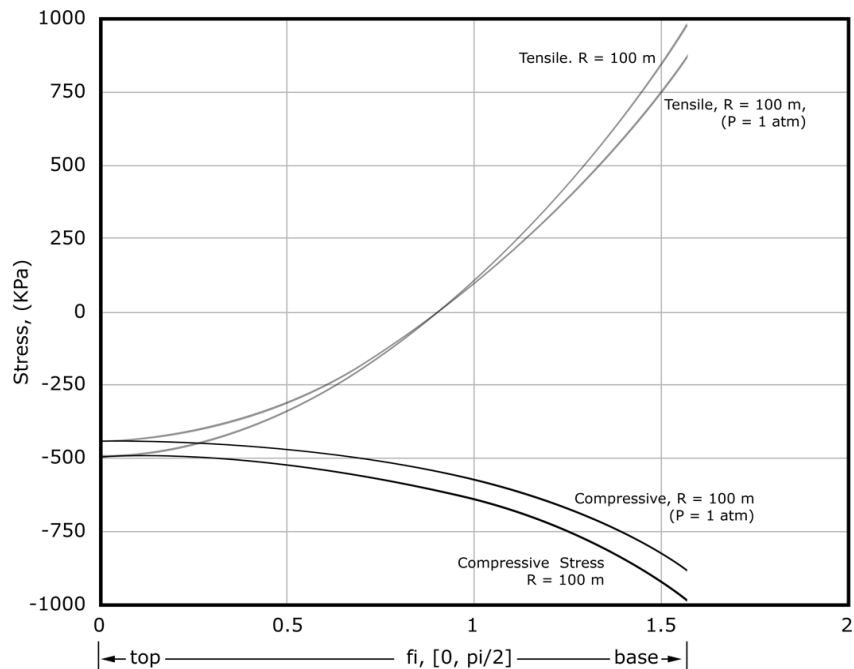

*Fig. 5: Compression and tensile stress in a 100 m radius hemispherical ice dome on Earth, with and without internal pressure.*

## 2.6  Benefits

**Large Pressurized Habitat:** a large pressurized living space would present an enormous advantage because it can revolutionize working efficiency and living conditions facilitating food growing on large scale and aiding development of industries like tourism. This might be a deciding factor in making colonization of outer space feasible and profitable.

**Radiation protection:** water has relatively good radiation shielding properties. The halving thickness of water for gamma rays is approximately $t_h$ = 14 cm [8], which means that a 1 meter wall of water will provide a decrease in gamma ray radiation of approximately $1/2^7$, effectively reducing the levels of radiation to 0.0078 of its original strength.

This means that colonists have the possibility to work under the dome almost indefinitely compared to a few hours a day without protection.

**Meteoroid protection:** space dwellers would be relatively well protected from small meteoroids falling onto the surface. Due the shape of the dome and the strength of the ice on the exterior wall, a large fraction of these incoming projectiles will ricochet causing only surface damage to the dome and the ETFE cover.

These defects can be repaired by injecting liquid water in the crack under pressure which will fuse with the surrounding ice restoring its original strength and the ETFE can be repaired by thermal patching [3].

In case a meteoroid punctures the dome, causing it to vent atmosphere, the atmospheric moisture from the inside could plug the whole as it condenses/deposits due to volume expansion and temperature drop in contact with the outer layers of the dome. If this doesn't happen spraying liquid water into the whole can seal the leak should this be of reasonable size.

**Recyclable material:** even though very large quantities of water are needed to construct domes, the material would be 100% recyclable. If a dome is scrapped due to structural fatigue or for any other reason, and a new one is being built nearby, the procedure can be reversed, by slowly sublimating the dome from inside out in low pressure conditions, and the water gas can be pumped to the nearby dome and used for construction. It could also be used for consumption, as it would be pure water which passed an evaporation process.

## 2.7 Shortcomings

**Construction related:** due to the physical properties of water, the method cannot be applied anywhere. If the pressure gets above 600 Pa, water will first melt into liquid state rather than going directly into gas, therefore further heating will be necessary to transform it into gas, or other dispersing mechanisms have to be considered such as fine spraying, both of which involve additional equipment.

With vacuum on the outside, the enormous force exerted by the one atmosphere pressure on the inside surface of the dome, will tend to lift dome. While this is good news because it aids the structural resistance by reducing stress caused by weight, for lower gravity conditions, such as on the moon or asteroids, the dome has to be specifically designed with thicker and/or higher shell to increase its weight just to keep it on the ground.

**Material Related:** Although outside temperature is suitable for maintaining an ice dome structurally stable, inside temperature which is suitable for plant and animal life, is considerably above the melting point of water. This can cause the structure to melt from the inside out.

In the long term, it would probably become feasible to transport equipment on site which can be used to create insulating materials from local raw material, similar to basaltic rock wool or even more elaborate ones like aerogels which can be used to insulate the structure from the inside and thus maintain a comfortable temperature.

In the short term, if insulation is not feasible, the inside temperature can be kept slightly below the melting point of water at atmospheric pressure, but still in the tolerable zone for people to work outside suitable dressed: something in the range of -3°C, -5°C. Although at these temperatures the strength of ice would be below the comfort zone on the inside, the thickness of the dome will ensure that enough of the structure would be within adequate temperature to keep the structure intact (at -8°, ice is already strong enough as the calculations in this paper show).

Even though this is less than ideal, lightweight structures inside the dome can sustain temperatures that are adequate for growing plants or living quarters.

## 2.8 Case Studies

From the structural consideration section and (Table 2) it can be observed that even though the compressive and tensile strength of ice are weak compared to regular construction materials, they are still somewhat higher (although only just) than the stress created in the shell of a one meter thick, two hundred meter wide dome on Earth. Earth is a good example to emphasize factor of safety for the building under its own weight because it is bigger than any other possible destination in space where such dome can be constructed, hence it will yield the lowest factor of safety of all.

On the Moon, where the gravitational acceleration is $g_m$ = 1.63 m/s$^2$ (only 16.7% that of Earth), this stress decreases to very low levels even for very large domes. Under these circumstances a 600 m wide ice dome on the moon, would need to bear a maximum compressive and tensile stress of approximately 500 KPa, which gives the construction a factor of safety, FoS = 1370/500 = 2.74 (Table 2), even at relatively high temperatures for ice ( -8°C), which is greater than 2.0, the standard factor of safety for buildings. Fig. 6 shows the evolution of the forces in the dome at various radii.

With atmospheric pressure inside, this stress would be reduced even further to approximately 200 KPa with a FoS = 6.85, for the 600 m diameter dome, the widest considered in the figure 6.

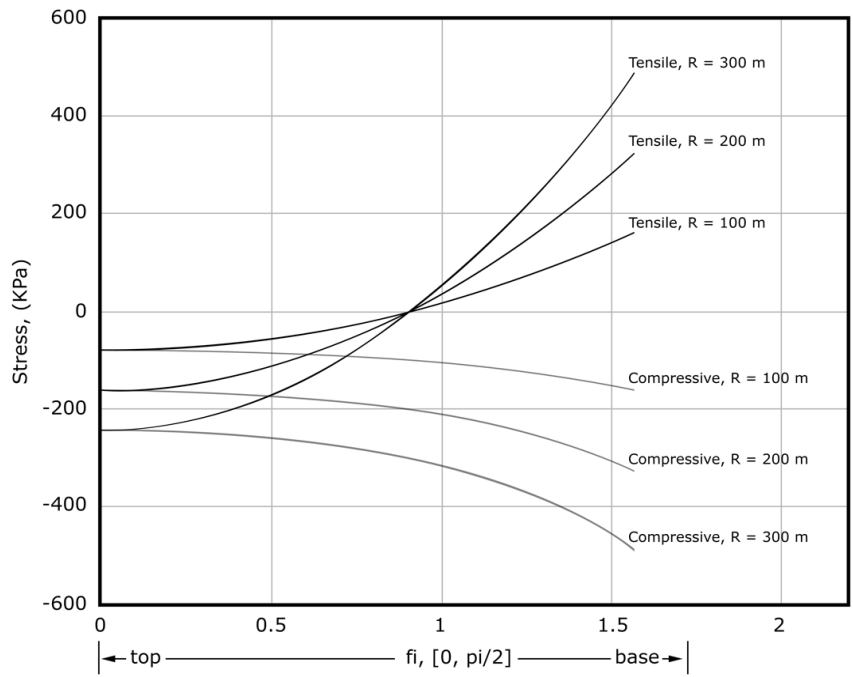

*Fig. 6: Evolution of stress in the dome on the Moon for various dome radii*

### 2.8.1 A Practical Example

Although it is interesting to see that such gigantic domes of ice are possible with regards to the stress inside the construction (a 600m diameter dome would cover a surface of more than 280,000 m²), for practical purposes a somewhat smaller dome will be considered:

- R = 50.0 m
- Max height = 15.0 (not a complete hemisphere)
- t = 4.00 m (The thick shell is needed to counteract the huge force produced by the inside atmosphere)
- r = 35.71 m
- $g_m$ = 1.63 m/s²
- Shell Volume = 15,716.34 m³
- Usable Area = 3,158.38 m²
- Foil Mass = 392.31 Kg
- Inflating Pressure = 7.2*10$^{-6}$ Kpa
- Heating Power = 39.5 KW
- Dome Weight = 25,617.64 KN
- Lift Force = 24,398.47 KN

- Residual Dome Weight = 1,219.17 KN
- gamma ray radiation atenuation = $2^{-28}$
- Stress: See Fig. 7 for stress evolutions in the shell. It can be seen that even the highest stress (in absolute values) is less than $S_{max}$ = 80 Kpa. This yields a huge factor of safety for the construction of FoS = 16.8, eight times that of standard constructions on Earth.

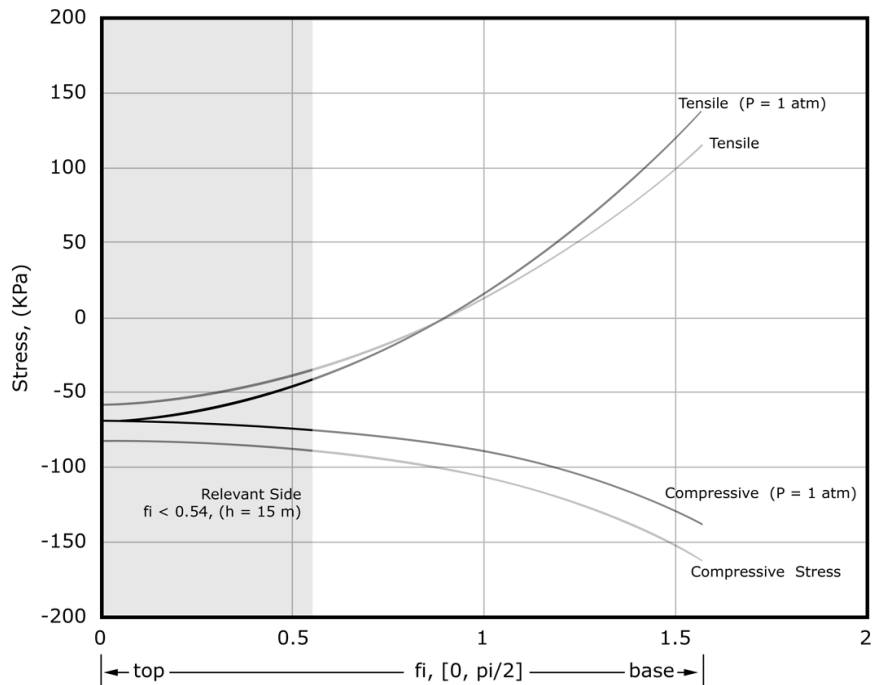

*Fig. 7: Tensile and compressive stress with and without atmosphere inside. Only the side to the left that is grayed (<~0.55) matters as the dome is not complete hemisphere.*

# 3 Conclusion

Although ice is not a usual construction material here on Earth, where temperature fluctuations make it unstable on the long term, in the vacuum of the Moon or other atmosphere-less, cold bodies it could prove an excellent construction material due to its properties which make it easy to handle.

Preliminary calculations show that for medium and large domes made of ice very good safety factors can be achieved even at high temperatures for ice (-8°C), whereas the conditions at the outer side of the shell in space would be closer to -200°C, depending on the interior insulation and interior temperature. At these values the endurance of ice is many times higher.

Long term stress tests would need to be conducted to determine problems that may occur due to plasticity of ice, which may allow for the deformation of the dome over a long period of time. Although current tests [2] show that forces would need to be much larger to cause significant deformation, compared to those that arise in the dome in the current calcualtions, these tests have not been conducted over long and very long periods of time (months, years).

Without a doubt, the possibility of large protected habitable areas would completely change the way we approach space exploration and colonization. It could be the difference between impossible and probable. This very simple method of construction could provide just that within the boundaries of technology that we already possess.

# 5 Tables

| | | |
|---|---|---|
| Temperature range of -254°C to +165°C (330°F). | | |
| very high degree of light transmission | | |
| good mechanical properties | | |
| broad chemical resistance | | |
| low flammability | | |
| very low surface energy | | |
| good stress-crack resistance | | |
| can easily be thermoformed: welded, fabricated, laminated | | |
| excellent dimensional stability during thermoforming | | |
| Tensile strength | 48,000 (KPa) | |
| Specific mass | 1771 (Kg/m$^3$) [1] | |
| Specific Mass per area for various thicknesses, | 0.0127 (mm) | 0.0225 (Kg/m$^2$) |
| | 0.0254 (mm) | 0.045 (Kg/m$^2$) |
| | 0.127 (mm) | 0.225 (Kg/m$^2$) |
| | 0.508 (mm) | 0.9 (Kg/m$^2$) |
| | | |

*Table 1: Mechanical properties of ETFE [3].*

*Ethylene tetrafluoroethylene is a Fluorine based plastic designed to have high corrosion and strength over a wide range of temperatures, properties that are ideal for use in space. As the method involves thin sheets of material, rather than volumes, the specific weight per area of material will be given for various thicknesses in the following table. The 0.0254 value is highlighted because it will be used in the case studies below.*

*A 20 nm thick low-e coating using silver film [9] will add to this 0.000054 Kg/m$^2$, or 0.5 Kg / 100,000 m$^2$, which will be ignored in the following calculations due to the extremely small value relative to that of the rest of the material.*

| | |
|---|---|
| Specific mass | 1000 (Kg/m$^3$) |
| Crushing strength at -5 ($^\circ$C) | 1,520 (Kpa) [2] |
| Crushing strength at -20 ($^\circ$C) | 2,720 (Kpa) [2] |
| Crushing strength at -40 ($^\circ$C) | 4,240 (Kpa) [2] |
| Crushing strength at -60 ($^\circ$C) | 5,360 (Kpa) [2] |
| Tensile Strength -8 ($^\circ$C) | 1,370 (Kpa) [2] |

*Table 2: Mechanical properties of water ice [2]. Specific mass, crushing and tensile strength [2] Specific mass of water will vary with the temperature but for convenience we will approximate it to specified value. For the tensile strength of ice only a few determinations of the tensile strength of ice are available (Romanowicz and Honigmann; Bernstein) [2], but it is enough to get a preliminary idea regarding the possibility of such structures.*